\documentclass[11p]{article}
\usepackage[dvips]{graphicx}
\usepackage{amsmath,amssymb,bm, ascmac}
\newcommand{\up}{{\rm up}}
\newtheorem{proposition}{Proposition}[section]
\newtheorem{theorem}{Theorem}[section]
\title{Ultradiscrete Permanent Solution to the Ultradiscrete KP Equation}
\author{
Hidetomo Nagai\thanks{ Department of Mathematics, Tokai University, 4-1-1, Kitakaname, Hiratsuka, Kanagawa, 259-1292, Japan. E-mail: hdnagai@tokai-u.jp}
}
\date{\today}
\begin{document}

\maketitle 

\begin{abstract}
We propose an ultradiscrete permanent solution to the ultradiscrete KP equation.  The ultradiscrete permanent is an ultradiscrete analogue of the usual permanent.  The elements on this ultradiscrete permanent solution are required some additional relations other than the ultradiscrete dispersion relation.  We confirm the solution satisfying these relations and propose some explicit examples of the solution.  
\end{abstract}
\section{Introduction}
Soliton equations are known as the ones possessing exact solutions, infinite conserved quantities and so on.  These solutions are generally expressed by the determinant in the form such as Wronskian, Grammian.  For such types of these solutions, its equations are transformed into the identity of determinants of which elements obey the dispersion relations.  One of the most fundamental soliton equations is the discrete KP equation\cite{Hirota1, Miwa}, which is expressed by 
\begin{equation}  \label{dKP}
\begin{aligned}
  &A_1(A_2-A_3)T(l+1, m, n)T (l, m+1, n+1) \\ 
  +&A_2(A_3-A_1)T(l, m+1, n)T (l+1, m, n+1) \\ 
  +&A_3(A_1-A_2)T (l, m, n+1)T (l+1, m+1, n)=0,   
\end{aligned}
\end{equation}
where $l$, $m$, $n$ are independent variables and $A_1$, $A_2$, $A_3$ are arbitrary parameters.  It is well known the discrete KP equation (\ref{dKP}) admits determinant solution\cite{Ohta}.  
\begin{equation}  \label{dKP sol}
  T(l, m, n) = \det
\begin{bmatrix}
  \varPhi _1(0) & \varPhi_1(1) & \cdots & \varPhi_1(N-1) \\
  \varPhi_2(0) & \varPhi_2(1) & \cdots & \varPhi_2(N-1) \\
  \vdots  & \vdots  & \ddots & \vdots  \\
  \varPhi_N(0) & \varPhi_N(1) & \cdots & \varPhi_N(N-1) 
\end{bmatrix},
\end{equation}
where $\varPhi_i(s)=\varPhi(l, m, n, s)$ satisfies the following dispersion relations,  
\begin{equation}  \label{discrete condition1}
\begin{aligned}
  &\varPhi_i(l+1, m, n, s)=  \varPhi_i(l, m, n, s)+A_1\varPhi_i(l, m, n, s+1), \\
  &\varPhi_i(l, m+1, n, s)=  \varPhi_i(l, m, n, s)+A_2\varPhi_i(l, m, n, s+1), \\
  &\varPhi_i(l, m, n+1, s)=  \varPhi_i(l, m, n, s)+A_3\varPhi_i(l, m, n, s+1),  
\end{aligned}
\end{equation}
for $i=1, 2, \dots , N$.  For example, the soliton solution can be obtained by setting 
\begin{equation}  \label{discrete soliton}
\begin{aligned}
  \varPhi_i(l, m, n, s)= &P_i^s(1+A_1P_i)^l(1+A_2P_i)^m(1+A_3P_i)^nC_i\\
    &+(-1)^{i+1}P_i^{-s}(1+A_1P_i)^{-l}(1+A_2P_i)^{-m}(1+A_3P_i)^{-n}C_i', 
\end{aligned}
\end{equation}
where $P_i$ and $C_i$, $C_i'$ are arbitrary parameters.  In fact, the function (\ref{discrete soliton}) satisfies (\ref{discrete condition1}). In addition, another solution also can be obtained by setting 
\begin{equation}  \label{det sol2}
\begin{bmatrix}
  \varPhi_1(s)\\
  \varPhi_2(s)\\
  \vdots  \\
  \varPhi_N(s)\\
\end{bmatrix}
=
\begin{bmatrix}
  C_{11} & C_{12} & \ldots & C_{1M}\\
  C_{21} & C_{22} & \ldots & C_{2M}\\
  \vdots  & \vdots & \ddots & \vdots \\
  C_{N1} & C_{N2} & \ldots & C_{NM}
\end{bmatrix}
\begin{bmatrix}
  P_1^s(1+A_1P_1)^l(1+A_2P_1)^m(1+A_3P_1)^nC_1\\
  P_2^s(1+A_1P_2)^l(1+A_2P_2)^m(1+A_3P_2)^nC_2\\
  \vdots\\
  P_M^s(1+A_1P_M)^l(1+A_2P_M)^m(1+A_3P_M)^nC_M
\end{bmatrix} , 
\end{equation}
where $P_i$, $C_{ij}$ are arbitrary parameters and $M$ is any positive integer.  Note that the dispersion relations (\ref{discrete condition1}) are general conditions to the solution (\ref{dKP sol}), and the functions (\ref{discrete soliton}) and (\ref{det sol2}) are only the specific realization of (\ref{discrete condition1}).  \par 
Ultradiscretization is a limiting procedure\cite{Tokihiro}.  By applying the ultradiscretization to the discrete KP equation, we can obtain the ultradiscrete KP(uKP) equation as below.  Introducing transformations $T(l, m, n) = e^{\tau(l, m, n)/\varepsilon }$, $A_i= e^{-a_i/\varepsilon }$ with a positive parameter $\varepsilon $, then (\ref{dKP}) is expressed by 
\begin{equation}  
\begin{aligned}
  &e^{(\tau(l+1, m, n)+\tau(l, m+1, n+1)-a_1-a_2)/\varepsilon } +e^{(\tau(l, m+1, n)+\tau(l+1, m, n+1)-a_2-a_3)/\varepsilon }\\
  & +e^{(\tau(l, m, n+1)+\tau(l+1, m+1, n)-a_3-a_1)/\varepsilon }  \\
  =&e^{(\tau(l+1, m, n)+\tau(l, m+1, n+1)-a_1-a_3)/\varepsilon } +e^{(\tau(l, m+1, n)+\tau(l+1, m, n+1)-a_2-a_1)/\varepsilon }\\
  & +e^{(\tau(l, m, n+1)+\tau(l+1, m+1, n)-a_2-a_3)/\varepsilon }  .  
\end{aligned}
\end{equation}
Applying $\varepsilon \log $ both sides and taking a limit $\varepsilon \to +0$, we obtain the uKP equation.  
\begin{equation}  \label{uKP eq}
\begin{split}
  \max( &\tau(l+1, m, n) +\tau(l, m+1, n+1)-a_1-a_2, \\
  &\tau(l, m+1, n) +\tau(l+1, m, n+1)-a_2-a_3, \\
  & \tau(l, m, n+1)+\tau(l+1, m+1, n)-a_1-a_3) \\
  =\max( &\tau(l+1, m, n) +\tau(l, m+1, n+1)-a_1-a_3, \\
  &\tau(l, m+1, n) +\tau(l+1, m, n+1)-a_1-a_2, \\
  & \tau(l, m, n+1)+\tau(l+1, m+1, n)-a_2-a_3),    
\end{split}
\end{equation} 
by using a key formula 
\begin{equation}  \label{ultradiscrete formula}
  \lim _{\varepsilon \to +0} \varepsilon \log (e^{a_1/\varepsilon }+e^{a_2/\varepsilon }+\dots +e^{a_n/\varepsilon }) =\max (a_1, a_2, \dots , a_n).  
\end{equation}  
Equivalently, (\ref{uKP eq}) is rewritten as 
\begin{equation}
\begin{split}  \label{uKP eq2}
  &\tau(l, m+1, n) +\tau(l+1, m, n+1), \\
  =&\max \bigl( \tau(l+1, m, n) +\tau(l, m+1, n+1)-a_1+a_2, \tau(l, m, n+1)+\tau(l+1, m+1, n)\bigr) 
\end{split}
\end{equation} 
under the assumption $a_1\ge a_2\ge a_3$ without a loss of generality.  We also obtain a solution to (\ref{uKP eq2}) if we can ultradiscretize (\ref{dKP sol}) and (\ref{discrete condition1}) straightforwardly.  However, the expression of sum of positive terms is required in order to apply the key formula (\ref{ultradiscrete formula}).   For this reason it is difficult to ultradiscretize determinant solution directly.  To avoid this problem, we introduce ultradiscrete permanent\cite{uKdV}.  The ultradiscrete permanent(UP) of $N\times N$ matrix $A=[a_{ij}]$ is defined as 
\begin{equation} 
\up[A]\equiv \max_{\pi } (a_{1\pi _1}+a_{2\pi_2}+\dots +a_{N\pi_N}),   
\end{equation}
where $\pi=(\pi_1, \pi_2, \dots , \pi_N)$ is a set of all possible permutations of $\{1, 2, \dots , N \}$.  Note that the UP is directly defined by ultradiscretizing a permanent by using (\ref{ultradiscrete formula}).  We proposed an UP solution to (\ref{uKP eq2}) in the previous paper\cite{Nagai1}.  It is an ultradiscrete analogue of (\ref{dKP sol}) associated with (\ref{discrete soliton}).  Also soliton solutions to the ultradiscrete KdV equation, the ultradiscrete Toda equation, the ultradiscrete hungry Lotka-Volterra equation are expressed by the UP form\cite{uKdV, Nagai2, SNakamura}.  These facts suggest UP is an ultradiscrete analogue of determinant.  However, the above solutions are ultradiscrete analogues of the soliton solutions, and UP solution such as determinant solution associated with dispersion relations is not obtained yet.  \par
In this paper, we propose an UP solution to the uKP equation.  The solution is an ultradiscrete analogue of (\ref{dKP sol}) associated with (\ref{discrete condition1}).  Note that we shall impose some additional conditions on the elements of UP to satisfy the uKP equation.  Moreover, we show some explicit elements which are ultradiscrete analogues of (\ref{discrete soliton}) and (\ref{det sol2}), and confirm they obey these conditions.  This paper is consists on below.  First, we give some properties of ultradiscrete permanent in section 2.  In section 3, we give an UP solution to the uKP equation.  In section 4, we propose some explicit examples of the solution.  Finally, we give concluding remarks in section 5.  \\
\section{Properties of Ultradiscrete Permanent}
Let $N$ be a positive integer. Consider an UP of $N\times N$ matrix $A=[a_{ij}]$.  Then one can derive the following properties from the definition.     
\begin{equation}  \label{prop1}
  \up 
\begin{bmatrix}
  a_{11} & \ldots & a_{1i} & \ldots & a_{1j} & \ldots & a_{1N}\\
  \vdots & \ddots & \vdots  & \ddots & \vdots  & \ddots & \vdots \\
  a_{N1} & \ldots & a_{Ni} & \ldots & a_{Nj} & \ldots & a_{NN}
\end{bmatrix}=
  \up 
\begin{bmatrix}
  a_{11} & \ldots & a_{1j} & \ldots & a_{1i} & \ldots & a_{1N}\\
  \vdots & \ddots & \vdots  & \ddots & \vdots  & \ddots & \vdots \\
  a_{N1} & \ldots & a_{Nj} & \ldots & a_{Ni} & \ldots & a_{NN}
\end{bmatrix} , 
\end{equation}
\begin{equation}  \label{prop2}
c+  \up 
\begin{bmatrix}
  a_{11} & \ldots & a_{1i}  & \ldots & a_{1N}\\
  \vdots & \ddots & \vdots  & \ddots & \vdots \\
  a_{N1} & \ldots & a_{Ni} & \ldots & a_{NN}
\end{bmatrix}=
  \up 
\begin{bmatrix}
  a_{11} & \ldots &c+ a_{1i} & \ldots & a_{1N}\\
  \vdots & \ddots & \vdots   & \ddots & \vdots \\
  a_{N1} & \ldots &c+ a_{Ni} & \ldots & a_{NN}
\end{bmatrix} ,
\end{equation}
and 
\begin{equation}  \label{prop3}
\begin{aligned}
 &\up 
\begin{bmatrix}
  a_{11} & \ldots & \max( a_{1i}, b_1)  & \ldots & a_{1N}\\
  \vdots & \ddots & \vdots  & \ddots & \vdots \\
  a_{N1} & \ldots & \max( a_{Ni}, b_N) & \ldots & a_{NN}
\end{bmatrix}\\
 =&
\max\left( \up 
\begin{bmatrix}
  a_{11} & \ldots & a_{1i} & \ldots & a_{1N}\\
  \vdots & \ddots & \vdots   & \ddots & \vdots \\
  a_{N1} & \ldots & a_{Ni} & \ldots & a_{NN}
\end{bmatrix}, \quad 
\up 
\begin{bmatrix}
  a_{11} & \ldots & b_1 & \ldots & a_{1N}\\
  \vdots & \ddots & \vdots   & \ddots & \vdots \\
  a_{N1} & \ldots & b_N & \ldots & a_{NN}
\end{bmatrix}
\right)  
\end{aligned}
\end{equation}
for $i, j=1, 2, \dots , N$.  Moreover, we give two propositions.  
\begin{proposition} \label{proposition1}
Let $N$, $M$ be positive integers, and $B=[b_{ij}]$, $C=[c_{ij}]$ an $N \times M$ matrix and an $M \times N$ matrix respectively.  Define ultradiscrete product of $B$ and $C$ as 
\begin{equation}
  B\otimes C = [ \max_{1\le k\le M} (b_{ik}+c_{kj}) ]_{1\le i, j\le N}.   
\end{equation}
Then,  
\begin{equation}  \label{formula1}
  \up [B\otimes C] =  \max _{1\le j_1\le \dots \le j_N\le M} \left( \up[B] ^{1\dots N}_{j_1\dots j_N}+ \up[C] _{1\dots N}^{j_1\dots j_N} \right) 
\end{equation}
holds.  Here $\up[B]^{1\dots N}_{j_1\dots j_N}$ denotes the minor UP of the matrix whose rows and columns are the those of $B$ at $1\dots N$ and $j_1\dots j_N$ respectively. 
\end{proposition}
\begin{proposition}  \label{proposition2}
Let $\bm a$, $\bm b$, $\bm c$ be arbitrary $N$-dimensional vectors.  Define $N\times N$ matrix $D_{l, m, n}$ as 
\begin{equation*}
  D_{l , m, n}= [\bm a \cdots \bm a  \ \bm b \cdots \bm b  \ \bm c \cdots \bm c]  ,
\end{equation*}
where $l$, $m$, $n$ denote numbers of the columns of $\bm a$, $\bm b$, $\bm c$ respectively.  If \begin{equation*}
  l+m+n=l'+m'+n'=N, \quad 0\le l<l', \quad  0\le m'<m , 
\end{equation*}
then 
\begin{equation}  \label{formula2}
\begin{aligned}
\up [D_{l, m, n}] +\up [D_{l' , m' , n' }] \le \up [D_{l+1 , m-1 , n}] +\up [D_{l'-1 , m'+1 , n' }] 
\end{aligned}
\end{equation}
holds.  
\end{proposition}
Proofs of Propositions are given in appendix.  
\section{UP Solution to the Ultradiscrete KP Equation}
In this section, we give the following theorem.  
\begin{theorem}  \label{theorem3.1}
Let $\tau (l, m, n)$ be 
\begin{equation}  \label{uKP sol}
  \tau(l, m, n) = \up
\begin{bmatrix}
  \varphi _1(0) & \varphi_1(1) & \cdots & \varphi_1(N-1) \\
  \varphi_2(0) & \varphi_2(1) & \cdots & \varphi_2(N-1) \\
  \vdots  & \vdots  & \ddots & \vdots  \\
  \varphi_N(0) & \varphi_N(1) & \cdots & \varphi_N(N-1) 
\end{bmatrix},
\end{equation}
where $\varphi_i(s)=\varphi_i(l, m, n, s)$ is a function depends on $l$, $m$, $n$ and $s$.  Suppose $\varphi_i(l, m, n, s)$ satisfies the following conditions.  
\begin{equation}  \label{condition1}
\begin{aligned}
  &\varphi_i(l+1, m, n, s)= \max( \varphi_i(l, m, n, s), \varphi_i(l, m, n, s+1)-a_1), \\
  &\varphi_i(l, m+1, n, s)= \max( \varphi_i(l, m, n, s), \varphi_i(l, m, n, s+1)-a_2), \\
  &\varphi_i(l, m, n+1, s)= \max( \varphi_i(l, m, n, s), \varphi_i(l, m, n, s+1)-a_3) ,
\end{aligned}
\end{equation}
\begin{equation}  \label{condition2}
  \varphi_{i_1}(s)+ \varphi_{i_2}(s)\le  \max\bigl( \varphi_{i_1}(s-1)+ \varphi_{i_2}(s+1),  \varphi_{i_2}(s-1)+ \varphi_{i_1}(s+1)\bigr) 
\end{equation}
for $1\le i, i_1, i_2\le N$, and   
\begin{equation}  \label{condition3}
\begin{aligned}
  &\up[\bm \varphi(0) \cdots \widehat{\bm \varphi(k_2)}\cdots \bm \varphi (N)]+ \up[\bm \varphi(0)\cdots \widehat{\bm \varphi(k_1)}\cdots \widehat{\bm \varphi(k_3)}\cdots  \bm \varphi (N+1)] \\
  =\max \Bigl( 
  &\up[\bm \varphi(0) \cdots \widehat{\bm \varphi(k_3)}\cdots \bm \varphi (N)]+ \up[\bm \varphi(0)\cdots \widehat{\bm \varphi(k_1)}\cdots \widehat{\bm \varphi(k_2)}\cdots  \bm \varphi (N+1)], \\
  &\up[\bm \varphi(0) \cdots \widehat{\bm \varphi(k_1)}\cdots \bm \varphi (N)]+ \up[\bm \varphi(0)\cdots \widehat{\bm \varphi(k_2)}\cdots \widehat{\bm \varphi(k_3)}\cdots  \bm \varphi (N+1)]\Bigr) , 
\end{aligned} 
\end{equation}
for $0\le k_1<k_2<k_3\le N$.  Here $\bm \varphi(s)$ denotes 
\begin{equation*}
  \bm \varphi(s) = 
\begin{bmatrix}
  \varphi_1(s) \\  \varphi_2(s) \\  \vdots \\  \varphi_N(s)  
\end{bmatrix}
\end{equation*}
and the symbol $\widehat {\bm \varphi(s)}$ means that $\bm \varphi(s)$ is omitted.  Then $\tau(l, m, n)$ satisfies (\ref{uKP eq2}).  
\end{theorem}
Note that (\ref{condition1}) is ultradiscrete analogue of (\ref{discrete condition1}).  Both (\ref{condition2}) and (\ref{condition3}) are additional conditions to satisfy the uKP equation (\ref{uKP eq2}).  Theorem \ref{theorem3.1} is proved by the similar manner given in \cite{Nagai1}.  Thus we show the outline in the case of $N=2$.  \par 
Using condition (\ref{condition1}) and the properties (\ref{prop2}), (\ref{prop3}), we can expand $\tau(l+1, m, n)$ as following.  
\begin{equation}
\begin{aligned}
  &\tau (l+1, m, n) \\
  =& \up 
\begin{bmatrix}
  \varphi_1(l+1, m, n, 0) & \varphi_1(l+1, m, n, 1) \\
  \varphi_2(l+1, m, n, 0) & \varphi_2(l+1, m, n, 1) 
\end{bmatrix} \\
  =& 
\up
\begin{bmatrix}
  \max( \varphi_1(0), \varphi_1(1)-a_1) & \max( \varphi_1(1), \varphi_1(2)-a_1) \\
  \max( \varphi_2(0), \varphi_2(1)-a_1) & \max( \varphi_2(1), \varphi_2(2)-a_1) 
\end{bmatrix} \\
  =& \max\Bigl( 
\up
\begin{bmatrix}
  \varphi_1(0) & \varphi_1(1) \\
  \varphi_2(0) & \varphi_2(1) 
\end{bmatrix}, 
\up
\begin{bmatrix}
  \varphi_1(1) & \varphi_1(1) \\
  \varphi_2(1) & \varphi_2(1) 
\end{bmatrix}-a_1, 
\up
\begin{bmatrix}
  \varphi_1(0) & \varphi_1(2) \\
  \varphi_2(0) & \varphi_2(2) 
\end{bmatrix}-a_1, 
\up
\begin{bmatrix}
  \varphi_1(1) & \varphi_1(2) \\
  \varphi_2(1) & \varphi_2(2) 
\end{bmatrix}-2a_1 \Bigr).  
\end{aligned}
\end{equation}
Moreover, since condition (\ref{condition2}) gives the inequality  
\begin{equation}
  \up
\begin{bmatrix}
  \varphi_1(s) & \varphi_1(s) \\
  \varphi_2(s) & \varphi_2(s) 
\end{bmatrix}\le 
 \up
\begin{bmatrix}
  \varphi_1(s-1) & \varphi_1(s+1) \\
  \varphi_2(s-1) & \varphi_2(s+1) 
\end{bmatrix},  
\end{equation}
thus we obtain 
\begin{equation}
\begin{aligned}
  \tau (l+1, m, n)  = \max\Bigl( 
\up
\begin{bmatrix}
  \varphi_1(0) & \varphi_1(1) \\
  \varphi_2(0) & \varphi_2(1) 
\end{bmatrix}, 
\up
\begin{bmatrix}
  \varphi_1(0) & \varphi_1(2) \\
  \varphi_2(0) & \varphi_2(2) 
\end{bmatrix}-a_1, 
\up
\begin{bmatrix}
  \varphi_1(1) & \varphi_1(2) \\
  \varphi_2(1) & \varphi_2(2) 
\end{bmatrix}-2a_1 \Bigr).  
\end{aligned}
\end{equation}
We also obtain the similar expressions of $\tau (l, m+1, n)$, $\tau (l, m, n+1)$, $\tau (l+1, m+1, n)$, $\tau (l+1, m, n+1)$, $\tau (l, m+1, n+1)$.  Substituting these expressions into LHS of (\ref{uKP eq2}), we obtain    
\begin{equation}  \label{LHS of uKP2}
\begin{aligned}
  &\tau(l, m+1, n) +\tau(l+1, m, n+1)\\
   =& \max\bigl( \up[0 \ 1], \up[0 \ 2]-a_2, \up[1 \ 2]-2a_2 \bigr)\\
  +& \max\bigl( \up[0 \ 1], \up[0 \ 2]-a_3, \up[0 \ 3]-a_1-a_3, \up[1 \ 2]-2a_3, \up[1 \ 3]-a_1-2a_3, \up[2 \ 3]-2a_1-2a_3 \bigr).  
\end{aligned}
\end{equation}
Here $[i \ j]$ denotes 
\begin{equation}
  [i \ j] = 
\begin{bmatrix}
  \varphi_1(i) & \varphi_1(j) \\
  \varphi_2(i) & \varphi_2(j) 
\end{bmatrix} .  
\end{equation}
On the other hand, RHS of (\ref{uKP eq2}) is expressed by 
\begin{equation}  \label{RHS of uKP2}
\begin{aligned}
  &\max( \tau(l+1, m, n) +\tau(l, m+1, n+1)-a_1+a_2, \tau(l, m, n+1) +\tau(l+1, m+1, n)) \\
   = \max\Bigl( &\max\bigl( \up[0 \ 1]-a_1, \up[0 \ 2]-2a_1, \up[1 \ 2]-3a_1 \bigr) + \max\bigl( \up[0 \ 1]+a_2, \up[0 \ 2]+a_2-a_3, \\
  & \up[0 \ 3]-a_3, \up[1 \ 2]+a_2-2a_3, \up[1 \ 3]-2a_3, \up[2 \ 3]-a_2-2a_3 \bigr), \\
    &\max\bigl( \up[0 \ 1], \up[0 \ 2]-a_3, \up[1 \ 2]-2a_3 \bigr) + \max\bigl( \up[0 \ 1], \up[0 \ 2]-a_2, \\
  &\up[0 \ 3]-a_1-a_2, \up[1 \ 2]-2a_2,   \up[1 \ 3]-a_1-2a_2, \up[2 \ 3]-2a_1-2a_2 \bigr)\Bigr).
\end{aligned}
\end{equation}
We can check the arguments which have the same coefficients of $a_1$, $a_2$, $a_3$ in (\ref{LHS of uKP2}) and (\ref{RHS of uKP2}) correspond respectively due to condition (\ref{condition3}).  For example, the argument which has $-a_1-a_2-2a_3$ in (\ref{LHS of uKP2}) is $\up[0 \ 2]+ \up[1 \ 3]$.  On the other hand, that in (\ref{RHS of uKP2}) is $\max(\up[0 \ 1]+ \up[2 \ 3], \up[1 \ 2]+\up[0 \ 3]) $.  They correspond for (\ref{condition3}).  Thus we have proved Theorem \ref{theorem3.1}.   
\section{Exact Solutions to the Ultradiscrete KP equation}
We propose explicit functions which satisfy (\ref{condition1}), (\ref{condition2}) and (\ref{condition3}).
\begin{proposition}  \label{proposition4.1}
Define 
\begin{equation}  \label{UP sol1}
\begin{aligned}
 \varphi_i(s)=\varphi_i(l, m, n, s)=&  \max \Bigl( p_i s+\max (0, p_i-a_1)l+\max (0, p_i-a_2)m+\max (0, p_i-a_3)n+c_i,\\
 &-p_i s+\max (0, -p_i-a_1)l+\max (0, -p_i-a_2)m+\max (0, -p_i-a_3)n+c'_i\Bigr), 
\end{aligned}
\end{equation}
where $p_i$ and $c_i$, $c_i'$ are arbitrary parameters.  Then $\varphi_i(s)$ satisfies (\ref{condition1}), (\ref{condition2}) and (\ref{condition3}).  
\end{proposition}
Proposition \ref{proposition4.1} is proved in \cite{Nagai1}.  
\begin{proposition}  \label{proposition4.2}
Define 
\begin{equation}  \label{UP sol2}
\begin{bmatrix}
  \varphi_1(s)\\
  \varphi_2(s)\\
  \vdots\\
  \varphi_N(s)\\
\end{bmatrix}
=
\begin{bmatrix}
  c_{11} & c_{12} & \ldots & c_{1M}\\
  c_{21} & c_{22} & \ldots & c_{2M}\\
  \vdots  & \vdots & \ddots & \vdots \\
  c_{N1} & c_{N2} & \ldots & c_{NM}
\end{bmatrix}
\otimes 
\begin{bmatrix}
  \eta_1(l, m, n, s)\\
  \eta_2(l, m, n, s)\\
  \vdots\\
  \eta_M(l, m, n, s)
\end{bmatrix},   
\end{equation}
where  
\begin{equation}  
  \eta _j(l, m, n, s)=p_js+ \max(0, p_j-a_1)l+\max(0, p_j-a_2)m+\max(0, p_j-a_3)n  ,
\end{equation}
and $c_{ij}$, $p_j$ are arbitrary parameters.  Then $\varphi_i(s)$ satisfies the conditions (\ref{condition1}), (\ref{condition2}) for any integer $M$, and (\ref{condition3}) in the case of $M=1, 2, 3$.  
\end{proposition}
We shall prove Proposition \ref{proposition4.2}.  Note (\ref{UP sol2}) is expressed by 
\begin{equation}  \label{phi}
  \varphi_i(l, m,  n, s) =\max _{1\le j\le M}\bigl( c_{ij}+ \eta _j(l, m, n, s)  \bigr).   
\end{equation}
Let us check (\ref{condition1}) first.  We have
\begin{equation}
\begin{aligned}
  \varphi_i(l+1, m, n, s) =& \max _{1\le j\le M}\bigl( c_{ij}+ \eta _j(l+1, m, n, s) \bigr)\\
  =& \max _{1\le j\le M}\bigl( c_{ij}+ \eta _j(l, m, n, s)+\max(0, p_j-a_1) \bigr)\\
  =&\max _{1\le j\le M}\bigl(\max(  c_{ij}+ \eta_j(l, m, n, s),  c_{ij}+ \eta_j(l, m, n, s+1)-a_1)\bigr) \\
  =&\max\bigl( \max _{1\le j\le M}\bigl(  c_{ij}+ \eta_j(l, m, n, s)\bigr) , \max _{1\le j\le M}\bigl( c_{ij}+ \eta_j(l, m, n, s+1)-a_1  \bigr)\bigr) \\
  =& \max(\varphi_i(l, m, n, s), \varphi_i(l, m, n, s+1)-a_1) .  
\end{aligned}
\end{equation}
Hence, 
\begin{equation}
  \varphi_i(l+1, m, n, s) = \max(\varphi_i(l, m, n, s), \varphi_i(l, m, n, s+1)-a_1)\end{equation}
holds.  The other relations in (\ref{condition1}) are also proved by similar manner.  \par
Next we consider (\ref{condition2}).  We have 
\begin{equation}
\begin{aligned}
  \varphi_{i_1}(s)+\varphi_{i_2}(s)   =& \max _{1\le j_1\le M}\bigl( c_{i_1j_1}+ \eta_{j_1}(l, m, n, s) \bigr)+ \max _{1\le j_2\le M}\bigl( c_{i_2j_2}+ \eta_{j_2}(l, m, n, s) \bigr) \\
  =&  \max _{1\le j_1, j_2\le M}\bigl( c_{i_1j_1}+c_{i_2j_2}+ \eta_{j_1}(l, m, n, s)+\eta_{j_2}(l, m, n, s)  \bigr).  
\end{aligned}
\end{equation}
On the other hand, 
\begin{equation}
\begin{aligned}
  &\max( \varphi_{i_1}(s-1)+\varphi_{i_2}(s+1), \varphi_{i_2}(s-1)+\varphi_{i_1}(s+1)) \\
  =&\max \bigl(  \max _{1\le j_1\le M}\bigl( c_{i_1j_1}+ \eta_{j_1}(l, m, n, s)-p_{j_1} \bigr)+  \max _{1\le j_2\le M}\bigl( c_{i_2j_2}+ \eta_{j_2}(l, m, n, s)+p_{j_2}  \bigr), \\
  &\qquad  \max _{1\le j_2\le M}\bigl( c_{i_2j_2}+ \eta_{j_2}(l, m, n, s)-p_{j_2} \bigr)+  \max _{1\le j_1\le M}\bigl( c_{i_1j_1}+ \eta_{j_1}(l, m, n, s)+p_{j_1}  \bigr) \bigr)\\
  =&\max \bigl(\max _{1\le j_1,j_2\le M}\bigl( c_{i_1j_1}+c_{i_2j_2}+ \eta_{j_1}(l, m, n, s)+\eta_{j_2}(l, m, n, s)-p_{j_1}+p_{j_2} \bigr), \\
  &\qquad  \max _{1\le j_1,j_2\le M}\bigl( c_{i_1j_1}+c_{i_2j_2}+ \eta_{j_1}(l, m, n, s)+\eta_{j_2}(l, m, n, s)+p_{j_1}-p_{j_2} \bigr)\bigr)\\
  =&\max _{1\le j_1,j_2\le M}\bigl( c_{i_1j_1}+c_{i_2j_2}+ \eta_{j_1}(l, m, n, s)+\eta_{j_2}(l, m, n, s) +\max(-p_{j_1}+p_{j_2}, p_{j_1}-p_{j_2})\bigr)\\
  =&\max _{1\le j_1,j_2\le M}\bigl( c_{i_1j_1}+c_{i_2j_2}+ \eta_{j_1}(l, m, n, s)+\eta_{j_2}(l, m, n, s) +|p_{j_1}-p_{j_2}|\bigr).
\end{aligned}
\end{equation}
Thus, (\ref{condition2}) also holds.  \par
Finally, let us consider (\ref{condition3}).  It holds if we prove an inequality
\begin{equation}  \label{ineq}
\begin{aligned}
 &\up[\bm \varphi(0) \cdots \widehat{\bm \varphi(k_2)}\cdots \bm \varphi (N)]+ \up[\bm \varphi(0)\cdots \widehat{\bm \varphi(k_1)}\cdots \widehat{\bm \varphi(k_3)}\cdots  \bm \varphi (N+1)] \\
 \ge &\up[\bm \varphi(0) \cdots \widehat{\bm \varphi(k_1)}\cdots \bm \varphi (N)]+ \up[\bm \varphi(0)\cdots \widehat{\bm \varphi(k_2)}\cdots \widehat{\bm \varphi(k_3)}\cdots  \bm \varphi (N+1)]
\end{aligned}
\end{equation}
since an identity 
\begin{equation}
\begin{aligned}
 \max\Bigl(
 &\up[\bm a_0 \cdots \widehat{\bm a_{k_2}}\cdots \bm a_N]+ \up[\bm a_0\cdots \widehat{\bm a_{k_1}}\cdots \widehat{\bm a_{k_3}}\cdots  \bm a_{N+1}] \\
  &\up[\bm a_0 \cdots \widehat{\bm a_{k_1}}\cdots \bm a_N]+ \up[\bm a_0\cdots \widehat{\bm a_{k_2}}\cdots \widehat{\bm a_{k_3}}\cdots  \bm a_{N+1}]\Bigr)\\
  =\max \Bigl( 
  &\up[\bm a_0 \cdots \widehat{\bm a_{k_3}}\cdots \bm a_N]+ \up[\bm a_0\cdots \widehat{\bm a_{k_1}}\cdots \widehat{\bm a_{k_2}}\cdots  \bm a_{N+1}] \\
  &\up[\bm a_0 \cdots \widehat{\bm a_{k_1}}\cdots \bm a_N]+ \up[\bm a_0\cdots \widehat{\bm a_{k_2}}\cdots \widehat{\bm a_{k_3}}\cdots  \bm a_{N+1}]\Bigr) 
\end{aligned}
\end{equation}
holds for any integers $0\le k_1<k_2<k_3\le N$ and any $N$-dimensional vectors $\bm a_j$\cite{Nagai1}.  For simplicity, we prove (\ref{ineq}) in the case of $N=2$, $M=3$ in this section.  A proof in the case of general $N$ and $M=1, 2, 3$ is given in appendix.  Set $N=2$, $M=3$, then $k_1$, $k_2$, $k_3$ are determined as $1$, $2$, $3$ uniquely and (\ref{ineq}) is reduced into 
\begin{equation} \label{ineq for n=2}
 \up[\bm \varphi(0) \ \bm \varphi (2)]+ \up[\bm \varphi(1) \ \bm \varphi (3)]  \ge \up[\bm \varphi(1) \ \bm \varphi (2)]+ \up[\bm \varphi(0) \ \bm \varphi (3)].     
\end{equation}
Note the inequality (\ref{ineq}) does not depend on $l$, $m$, $n$, thus we can assume 
\begin{equation}  
  \varphi_i(s) = \max_{1\le j\le 3}( c_{ij}+p_js)  
\end{equation}
without a loss of generality.  Then $\up[\bm \varphi(0) \ \bm \varphi (2)]$ is expressed by 
\begin{equation*}
\begin{aligned}
  &\up[\bm \varphi(0) \ \bm \varphi (2)]
  = \up 
\begin{bmatrix}
\begin{bmatrix}
  c_{11} & c_{12} & c_{13} \\
  c_{21} & c_{22} & c_{23} \\
\end{bmatrix}
\otimes 
\begin{bmatrix}
  0 & 2p_1 \\
  0 & 2p_2 \\
  0 & 2p_3 
\end{bmatrix}  
\end{bmatrix}.
\end{aligned}
\end{equation*}
We can also assume 
\begin{equation}  \label{ineq of p}
  p_1\le p_2\le p_3.  
\end{equation}
Thus, Using (\ref{formula1}) and (\ref{ineq of p}),  $\up[\bm \varphi(0) \ \bm \varphi (2)]$  can be expanded as 
\begin{equation*}
\begin{aligned}
  \up[\bm \varphi(0) \ \bm \varphi (2)]
  =& \max_{1\le i_1\le i_2\le 3} \Biggl(  \up 
\begin{bmatrix}
  c_{1i_1} & c_{1i_2} \\
  c_{2i_1} & c_{2i_2}  \\
\end{bmatrix}
+ \up
\begin{bmatrix}
  0 & 2p_{i_1} \\
  0 & 2p_{i_2} 
\end{bmatrix}
\Biggr)\\
  =&   \max_{1\le i_1\le i_2\le 3} \bigl(  \up[C_{i_1i_2}] +2 p_{i_2}\bigr)
\end{aligned}
\end{equation*}
where  $C_{i_1i_2}$ denotes
\begin{equation*}
\begin{aligned}
  C_{i_1i_2}= 
\begin{bmatrix}
  c_{1i_1} & c_{1i_2} \\
  c_{2i_1} & c_{2i_2} 
\end{bmatrix}.  
\end{aligned}
\end{equation*}
  By similar procedure, we obtain 
\begin{equation}  \label{LHS for n=2}
  \up[\bm \varphi(0) \ \bm \varphi (2)]+ \up[\bm \varphi(1) \ \bm \varphi (3)]= \max_{I, J}\bigl( \up[C_{i_1i_2}]+\up[C_{j_1j_2}]+ 2p_{i_2}+p_{j_1}+3p_{j_2} \bigr), 
\end{equation}
\begin{equation}  \label{RHS for n=2}
  \up[\bm \varphi(1) \ \bm \varphi (2)]+ \up[\bm \varphi(0) \ \bm \varphi (3)]  = \max_{I', J'}\bigl( \up[C_{i_1'i_2'}]+\up[C_{j_1'j_2'}]+ p_{i'_1}+2p_{i'_2}+3p_{j'_2} \bigr).   
\end{equation}
Here we denote $\max_{1\le i_1\le i_2\le 3, 1\le j_1\le j_2\le 3}$ as $\max_{I, J}$.  Hereafter we fix $i_1'$, $i_2'$, $j_1'$, $j_2'$.  Then we have
\begin{equation}
\begin{aligned}
  &\max_{I, J}\bigl( \up[C_{i_1i_2}]+\up[C_{j_1j_2}]+ 2p_{i_2}+p_{j_1}+3p_{j_2} \bigr) - (\up[C_{i_1'i_2'}]+\up[C_{j_1'j_2'}]+ p_{i'_1}+2p_{i'_2}+3p_{j'_2}) \\
  \ge &\max\bigl( \up[C_{i_1'i_2'}]+\up[C_{j_1'j_2'}] +2p_{i'_2}+p_{j'_1}+3p_{j'_2}, \up[C_{j_1'j_2'}]+\up[C_{i_1'i_2'}] +2p_{j'_2}+p_{i'_1}+3p_{i'_2} \bigr)\\
  &\quad   - (\up[C_{i_1'i_2'}]+\up[C_{j_1'j_2'}]+ p_{i'_1}+2p_{i'_2}+3p_{j'_2}) \\
  =&\max(p_{j'_1}-p_{i'_1}, -p_{j'_2}+p_{i'_2})  .  
\end{aligned}
\end{equation}
It takes a nonnegative value except in the case of $j'_1<i'_1\le i'_2<j'_2$, namely, $i'_1=i'_2=2$, $j'_1=1$, $j'_2=3$.  When $i'_1=i'_2=2$, $j_1'=1$, $j'_2=3$, we have  
\begin{equation}
\begin{aligned}
  &\max_{I, J}\bigl( \up[C_{i_1i_2}]+\up[C_{j_1j_2}]+ 2p_{i_2}+p_{j_1}+3p_{j_2} \bigr) - (\up[C_{22}]+\up[C_{13}]+p_2+2p_2+3p_3 ) \\
  \ge &\up[C_{12}]+\up[C_{23}] +2p_2+p_2+3p_3 - (\up[C_{22}]+\up[C_{13}]+p_2+2p_2+3p_3 ) \\
  =&\up[C_{12}]+\up[C_{23}]-\up[C_{22}]-\up[C_{13}].  
\end{aligned}
\end{equation}
It also takes a nonnegative value from Proposition \ref{proposition2}.  Thus, we have proved that (\ref{RHS for n=2}) is less than or equal to (\ref{LHS for n=2}).  Therefore (\ref{condition3}) holds.  \par
Note (\ref{ineq for n=2}) does not hold when $M\ge 4$.  This is because the term which has $\up[C_{23}]+\up[C_{14}]$ in (\ref{RHS for n=2}) may be greater than (\ref{LHS for n=2}).  
\section{Concluding Remarks}
In this paper, we show the UP defined (\ref{uKP sol}) under the conditions (\ref{condition1}), (\ref{condition2}), (\ref{condition3}) satisfies the uKP equation.  It is proved by using some properties of the ultradiscrete permanents.  Moreover we give some explicit solutions.  We may regard (\ref{UP sol1}) and (\ref{UP sol2}) as the ultradiscrete analogues of (\ref{discrete soliton}) and (\ref{det sol2}).  The uKP equation admits the UP solution with (\ref{UP sol2}) in the case of $M\le 3$ although the discrete KP equation does for any $M$.  To clarify these differences is one of the future problems.    
\section*{Acknowledgement}
The author is grateful to Professor Daisuke Takahashi for helpful advice.  The author also grateful Keisuke Mizuki for valuable discussions.
\appendix
\section{Proof of (\ref{formula1})}
In this appendix, we prove Proposition \ref{proposition1}.  From the definition, $\up [B\otimes C]$ is expressed by 
\begin{equation}   \label{app1}
  \up[B \otimes C] = \up 
\begin{bmatrix}
  \max_{1\le k\le M}(b_{1k}+c_{k1}) & \ldots  & \max_{1\le k\le M}(b_{1k}+c_{kN}) \\
  \vdots  & \ddots  & \vdots  \\
  \max_{1\le k\le M}(b_{Nk}+c_{k1}) & \ldots  & \max_{1\le k\le M}(b_{Nk}+c_{kN}) 
\end{bmatrix}.  
\end{equation}
Applying (\ref{prop2}) and (\ref{prop3}) to the first column, (\ref{app1}) is expanded as the maximum of $M$ UPs as below.  
\begin{equation}
\begin{aligned}
  &\up[B\otimes C]=\max\Biggl( \up 
\begin{bmatrix}
  b_{11}+c_{11} & \ldots  & \max_{1\le k\le M}(b_{1k}+c_{kN}) \\
  \vdots  & \ddots  & \vdots  \\
  b_{N1}+c_{11} & \ldots  & \max_{1\le k\le M}(b_{Nk}+c_{kN}) 
\end{bmatrix}, \\ 
&\up 
\begin{bmatrix}
  b_{12}+c_{21} & \ldots  & \max_{1\le k\le M}(b_{1k}+c_{kN}) \\
  \vdots  & \ddots  & \vdots  \\
  b_{N2}+c_{21} & \ldots  & \max_{1\le k\le M}(b_{Nk}+c_{kN}) 
\end{bmatrix}, \ \dots , \up 
\begin{bmatrix}
  b_{1M}+c_{M1} & \ldots  & \max_{1\le k\le M}(b_{1k}+c_{kN}) \\
  \vdots  & \ddots  & \vdots  \\
  b_{NM}+c_{M1} & \ldots  & \max_{1\le k\le M}(b_{Nk}+c_{kN}) 
\end{bmatrix}\Biggr) \\
 & = \max_{1\le k_1\le M} \Biggl( c_{k_11}+\up 
\begin{bmatrix}
  b_{1k_1} & \ldots  & \max_{1\le k\le M}(b_{1k}+c_{kN}) \\
  \vdots  & \ddots  & \vdots  \\
  b_{Nk_1} & \ldots  & \max_{1\le k\le M}(b_{Nk}+c_{kN}) 
\end{bmatrix}\Biggr) .
\end{aligned}
\end{equation}
Applying similar procedure to the other columns, we obtain 
\begin{equation}
\begin{aligned}
 \up[B\otimes C]=  \max _{1\le k_1, k_2, \dots , k_N\le M} \Biggl( \sum _{1\le i\le N} c_{k_ii} +\up 
\begin{bmatrix}
  b_{1k_1} & \ldots  & b_{1k_N} \\
  \vdots  & \ddots  & \vdots  \\
  b_{Nk_1} & \ldots  & b_{Nk_N} 
\end{bmatrix}\Biggr) .  
\end{aligned}
\end{equation}
It is equivalent to  
\begin{equation}
\begin{aligned}
 \up[B\otimes C]=&  \max_{1\le j_1\le \dots \le j_N\le M} \left( \max_{\pi'}\left( \sum _{1\le i\le N} c_{\pi' _ii} +\up 
\begin{bmatrix}
  b_{1\pi'_1} & \ldots  & b_{1\pi'_N} \\
  \vdots  & \ddots  & \vdots  \\
  b_{N\pi'_1} & \ldots  & b_{N\pi'_N} 
\end{bmatrix}\right) \right), 
\end{aligned}
\end{equation}
where $\pi'=(\pi'_1, \pi'_2, \dots , \pi'_N)$  is a set of all possible permutations of $\{j_1, j_2, \dots , j_N \}$.  In particular, the UP of the matrix whose columns are exchanged is the same as original one from (\ref{prop1}).  Therefore, we obtain 
\begin{equation}
\begin{aligned}
 \up[B\otimes C]=&
\max_{1\le j_1\le \dots \le j_N\le M} \left( \max_{\pi' }  \sum _{1\le i\le N} c_{\pi'_ii} +\up 
\begin{bmatrix}
  b_{1j_1} & \ldots  & b_{1j_N} \\
  \vdots  & \ddots  & \vdots  \\
  b_{Nj_1} & \ldots  & b_{Nj_N} 
\end{bmatrix}\right) \\
  =&\max_{1\le j_1\le \dots \le j_N\le M} \left( \up[C]^{j_1\dots j_N}_{1\dots N} +\up[B]^{1\dots N}_{j_1\dots j_N} \right).   
\end{aligned}
\end{equation}
Thus (\ref{formula1}) holds.  
\section{Proof of (\ref{formula2})} 
In this appendix, we express  
\begin{equation*}
  \bm a = \begin{bmatrix} a_1\\ a_2 \\ \vdots \\ a_N\end{bmatrix}, \quad 
  \bm b = \begin{bmatrix} b_1\\ b_2 \\ \vdots \\ b_N\end{bmatrix}, \quad 
  \bm c = \begin{bmatrix} c_1\\ c_2 \\ \vdots \\ c_N\end{bmatrix},
\end{equation*}
respectively.  Then $\up [D_{l , m, n}]+\up [D_{l' , m', n'}]$ is expressed by 
\begin{equation}  
\begin{aligned}
 \up [D_{l , m, n}] +\up [D_{l' , m' , n' }] =&  \max_{\pi , \pi' }\bigl( a_{\pi_1}+\dots + a_{\pi _l}+b_{\pi _{l+1}}+\dots +b_{\pi _{l+m}}+ c_{\pi_{l+m+1}}+\dots + c_{\pi _N} \\ 
  &+  a_{\pi'_1}+\dots + a_{\pi' _{l' }}+b_{\pi' _{l' +1}}+\dots +b_{\pi' _{l' +m'}}+ c_{\pi'_{l' +m' +1}}+\dots + c_{\pi' _N}\bigr),  
\end{aligned}
\end{equation}
namely, $\up [D_{l , m, n}]+\up [D_{l' , m', n'}]$ can be expressed by 
\begin{equation}  \label{argument}
\begin{aligned}
  & a_{\pi_1}+\dots + a_{\pi _l}+b_{\pi _{l+1}}+\dots +b_{\pi _{l+m}}+ c_{\pi_{l+m+1}}+\dots + c_{\pi _N} \\
  &+  a_{\pi'_1}+\dots + a_{\pi' _{l' }}+b_{\pi' _{l' +1}}+\dots +b_{\pi' _{l' +m'}}+ c_{\pi'_{l' +m' +1}}+\dots + c_{\pi' _N}  
\end{aligned}
\end{equation}
 for certain permutations $\pi $ and $\pi'$.  In particular, due to $m>m'$, there exists $j$ such that \begin{equation*}
\begin{aligned}
 j\in & \{ \pi_{l+1 }, \pi_{l +2 }, \dots , \pi_{l+m }\}\quad \text{and } \quad  j\in \{ \pi'_1 , \pi'_2 , \dots , \pi'_{l'}\} \cup \{ \pi'_{l'+m'+1} , \pi'_{l'+m'+2} , \dots , \pi'_N\}.     
\end{aligned}
\end{equation*}
First, let us consider in the case there exists $j_0$ such that 
\begin{equation}  \label{j0}
 j_0\in  \{ \pi_{l+1 }, \pi_{l +2 }, \dots , \pi_{l+m }\}\quad \text{and } \quad  j_0\in \{ \pi'_1 , \pi'_2 , \dots , \pi'_{l'}\}.  
\end{equation}
Then $\up [D_{l, m, n}] +\up [D_{l' , m' , n' }]$ can be expanded as
\begin{equation} \label{A-2-1}
\begin{aligned}
  \up [D_{l, m, n}] +\up [D_{l' , m' , n' }]   =& \up \left[ D_{l, m, n} \begin{bmatrix} j_0 \\ l+1 \end{bmatrix}\right]+b_{j_0} + \up \left[D_{l' , m' , n'} \begin{bmatrix} j_0 \\ 1 \end{bmatrix}\right]+a_{j_0} 
\end{aligned}
\end{equation}
where $D \begin{bmatrix} j \\ k \end{bmatrix}$ denotes the $(N-1)\times (N-1)$ matrix obtained by eliminating the $j$th row and the $k$th column from $D$.  On the other hand, $\up [D_{l+1 , m-1 , n}] +\up [D_{l'-1 , m'+1 , n' }]$ can be evaluated as 
\begin{equation}  \label{A-2-2}
\begin{aligned}
 \up [D_{l+1 , m-1 , n}] +\up [D_{l'-1 , m'+1 , n' }]  \ge &\up \left[D_{l+1 , m-1 , n}\begin{bmatrix} j_0 \\ 1 \end{bmatrix}\right]+a_{j_0} +\up \left[D_{l'-1 , m'+1 , n' }\begin{bmatrix} j_0 \\ l'  \end{bmatrix}\right]+b_{j_0}.  
\end{aligned}
\end{equation}
From (\ref{A-2-1}) and (\ref{A-2-2}), we obtain $\up [D_{l+1 , m-1 , n}] +\up [D_{l'-1 , m'+1 , n' }]\ge \up [D_{l , m , n}] +\up [D_{l' , m' , n' }]$ since 
\begin{equation}  
\begin{aligned}
  &\up \left[ D_{l, m, n} \begin{bmatrix} j_0 \\ l+1 \end{bmatrix}\right]= \up \left[D_{l+1 , m-1 , n}\begin{bmatrix} j_0 \\ 1 \end{bmatrix}\right] ,\quad \up \left[D_{l' , m' , n'} \begin{bmatrix} j_0 \\ 1 \end{bmatrix}\right] = 
\up \left[D_{l'-1 , m'+1 , n' }\begin{bmatrix} j_0 \\ l'  \end{bmatrix}\right]  
\end{aligned}
\end{equation}
hold.  \par
Next we consider the case there is no $j_0$ such that (\ref{j0}).  Then there exists $j_1$ such that 
\begin{equation*}
 j_1\in  \{ \pi_{l+1 }, \pi_{l +2 }, \dots , \pi_{l+m }\}\quad \text{and } \quad  j_1\in \{ \pi'_{l'+m'+1} , \pi'_{l'+m'+2} , \dots , \pi'_N\} .  
\end{equation*}
In addition, due to $l<l'$, there also exists $j_2$ such that 
\begin{equation*}
  j_2\in  \{ \pi_{l +m+1} , \pi_{l+m+2} , \dots , \pi_N \} \quad \text{and } \quad  j_2\in  \{ \pi'_1 , \pi'_2 , \dots , \pi'_{l'} \} .  
\end{equation*}
Thus,  $\up [D_{l, m, n}] +\up [D_{l' , m' , n' }]$ can be expanded as 
\begin{equation}  \label{A-2-3}
\begin{aligned}
  &\up [D_{l, m, n}] +\up [D_{l' , m' , n' }] \\
=& \up \left[D_{l, m , n} \begin{bmatrix} j_1 & j_2 \\ l+1 & l +m+1\end{bmatrix}\right]+b_{j_1}+c_{j_2} + \up \left[D_{l' , m' , n'} \begin{bmatrix}j_1 & j_2 \\ l'+m'+1  & 1 \end{bmatrix}\right]+c_{j_1}+a_{j_2} .
\end{aligned}
\end{equation}
On the other hand $\up [D_{l+1 , m-1 , n}] +\up [D_{l'-1 , m'+1 , n' }]$ can be evaluated as 
\begin{equation}  \label{A-2-4}
\begin{aligned}
 &\up [D_{l+1 , m-1 , n}] +\up [D_{l'-1 , m'+1 , n' }] \\
 \ge &\up \left[D_{l+1 , m-1 , n }\begin{bmatrix} j_1 & j_2 \\ l+m+1 & 1 \end{bmatrix}\right]+c_{j_1}+a_{j_2} +\up \left[ D_{l'-1 , m'+1 , n' }\begin{bmatrix} j_1 & j_2 \\ l' &  l'+m'+1   \end{bmatrix}\right]+b_{j_1}+c_{j_2}.  
\end{aligned}
\end{equation}
From (\ref{A-2-3}) and (\ref{A-2-4}), we obtain $\up [D_{l+1 , m-1 , n}] +\up [D_{l'-1 , m'+1 , n' }]\ge \up [D_{l , m , n}] +\up [D_{l' , m' , n' }]$ since 
\begin{equation}  
\begin{aligned}
  &\up \left[D_{l, m , n} \begin{bmatrix} j_1 & j_2 \\ l+1 & l +m+1\end{bmatrix}\right] = \up \left[D_{l+1 , m-1 , n }\begin{bmatrix} j_1 & j_2 \\ l+m+1 & 1 \end{bmatrix}\right]  ,  \\
  &\up \left[D_{l' , m' , n'} \begin{bmatrix}j_1 & j_2 \\ l'+m'+1  & 1 \end{bmatrix}\right]=\up \left[ D_{l'-1 , m'+1 , n' }\begin{bmatrix} j_1 & j_2 \\ l' &  l'+m'+1   \end{bmatrix}\right]  
\end{aligned}
\end{equation}
hold.  
\section{Proof of (\ref{ineq})}
We prove (\ref{ineq}) in the case of $M=1, 2, 3$ and any integer $N$.  We can assume
\begin{equation}  
  \varphi_i(s) = \max_{1\le j\le M}( c_{ij}+p_js)
\end{equation}
and 
\begin{equation}  \label{ineq of p2}
  p_1\le p_2\le \dots \le p_M
\end{equation}
without a loss of generality.  One can prove (\ref{ineq}) in the case of $M=1$.  In the case of $M=2$,  by adding $-\sum _{i=1}^N(c_{i1}+c_{i2})- \sum_{i=1}^N (p_1+p_2)i- \frac{(p_1+p_2)(N+1-k_1-k_2-k_3)}{2}$ to (\ref{ineq}), it is reduced into the inequality proved in \cite{Nagai1} since each $\varphi_i(s)$ can be rewritten as  
\begin{equation*}
  \varphi_i(s)-\frac{c_{i1}+c_{i2}+p_1s+p_2s}{2} = \frac{1}{2}|c_{i1}-c_{i2}+(p_1-p_2)s|.  
\end{equation*}
  Thus (\ref{ineq}) holds.  \par
We consider in the case of $M=3$.  Using (\ref{formula1}), $\up[\bm \varphi(0) \cdots \widehat{\bm \varphi(k_2)}\cdots \bm \varphi (N)]$ can be expanded as 
\begin{equation*}
\begin{aligned}
  &\up[\bm \varphi(0) \cdots \widehat{\bm \varphi(k_2)}\cdots \bm \varphi (N)]\\
  =& \up 
\begin{bmatrix}
\begin{bmatrix}
  c_{11} & c_{12} & c_{13} \\
  c_{21} & c_{22} & c_{23} \\
  \vdots  & \vdots  & \vdots  \\
  c_{N1} & c_{N2} & c_{N3} 
\end{bmatrix}
\otimes 
\begin{bmatrix}
  0 & p_1 & 2p_1 &  \ldots & \widehat{k_2p_1} & \ldots & Np_1\\
  0 & p_2 & 2p_2 &  \ldots & \widehat{k_2p_2} & \ldots & Np_2\\
  0 & p_3 & 2p_3 &  \ldots & \widehat{k_2p_3} & \ldots & Np_3
\end{bmatrix}
\end{bmatrix}\\
  =& \max_{1\le i_1\le \dots \le i_N\le 3} \Biggl(  \up 
\begin{bmatrix}
  c_{1i_1} & c_{1i_2} & \ldots  & c_{1i_N} \\
  c_{2i_1} & c_{2i_2} & \ldots  & c_{2i_N} \\
  \vdots  & \vdots &\ddots  & \vdots  \\
  c_{Ni_1} & c_{Ni_2} & \ldots  & c_{Ni_N} 
\end{bmatrix}
+ \up
\begin{bmatrix}
  0 & p_{i_1} & 2p_{i_1} &  \ldots & \widehat{k_2p_{i_1}} & \ldots & Np_{i_1}\\
  0 & p_{i_2} & 2p_{i_2} &  \ldots & \widehat{k_2p_{i_2}} & \ldots & Np_{i_2}\\
  \vdots  & \vdots & \vdots &  \ddots & \vdots  & \ddots & \vdots \\
  0 & p_{i_N} & 2p_{i_N} &  \ldots & \widehat{k_2p_{i_N}} & \ldots & Np_{i_N}
\end{bmatrix}
\Biggr)\\
 =& \max_{1\le i_1\le \dots \le i_N\le 3}\Biggl(  \up[C_{i_1i_2\dots i_N}]+ \sum_{l=1}^{k_2} (l-1) p_{i_l}+\sum_{l=k_2+1}^{N} l p_{i_l}\Biggr),  
\end{aligned}
\end{equation*}
where  $ C_{i_1i_2\dots i_N}$ denotes
\begin{equation*}
\begin{aligned}
  C_{i_1i_2\dots i_N}=  
\begin{bmatrix}
  c_{1i_1} & c_{1i_2} & \ldots  & c_{1i_N} \\
  c_{2i_1} & c_{2i_2} & \ldots  & c_{2i_N} \\
  \vdots  & \vdots &\ddots  & \vdots  \\
  c_{Ni_1} & c_{Ni_2} & \ldots  & c_{Ni_N} 
\end{bmatrix}.  
\end{aligned}
\end{equation*}
By similar procedure, we obtain 
\begin{equation}  \label{LHS}
\begin{aligned}
  &\up[\bm \varphi(0) \cdots \widehat{\bm \varphi(k_2)}\cdots \bm \varphi (N)]+ \up[\bm \varphi(0)\cdots \widehat{\bm \varphi(k_1)}\cdots \widehat{\bm \varphi(k_3)}\cdots  \bm \varphi (N+1)]\\
  =& \max_{I, J}\Bigl( \up[C_{i_1i_2\dots i_N}]+\up[C_{j_1j_2\dots j_N}]+ 
\sum_{l=1}^N l p_{i_l}+ \sum_{l=1}^N lp_{j_l}-\sum_{l=1}^{k_2} p_{i_l}-\sum_{l=1}^{k_1}  p_{j_l} +\sum_{l=k_3}^N  p_{j_l}\Bigr),   
\end{aligned}
\end{equation}
\begin{equation}  \label{RHS}
\begin{aligned}
  &\up[\bm \varphi(0) \cdots \widehat{\bm \varphi(k_1)}\cdots \bm \varphi (N)]+ \up[\bm \varphi(0)\cdots \widehat{\bm \varphi(k_2)}\cdots \widehat{\bm \varphi(k_3)}\cdots  \bm \varphi (N+1)]\\
  =& \max_{I', J'}\Bigl( \up[C_{i'_1i'_2\dots i'_N}]+\up[C_{j'_1j'_2\dots j'_N}]+ 
\sum_{l=1}^N l p_{i'_l}+ \sum_{l=1}^N lp_{j'_l}-\sum_{l=1}^{k_1} p_{i'_l}-\sum_{l=1}^{k_2}  p_{j'_l} +\sum_{l=k_3}^N  p_{j'_l}\Bigr).  
\end{aligned}
\end{equation}
Here we denote $\max_{1\le i_1\le \dots \le i_N\le 3, 1\le j_1\le \dots \le j_N\le 3}$ as $\max_{I, J}$ and $\sum _{l=m}^n$ is defined as $0$ when $m>n$.  Let us consider the argument of (\ref{RHS}):  
\begin{equation}  \label{argument of RHS}
  \up[C_{i'_1i'_2\dots i'_N}]+\up[C_{j'_1j'_2\dots j'_N}]+ \sum_{l=1}^N l p_{i'_l}+ \sum_{l=1}^N lp_{j'_l}-\sum_{l=1}^{k_1} p_{i'_l}-\sum_{l=1}^{k_2}  p_{j'_l} +\sum_{l=k_3}^N  p_{j'_l}.  
\end{equation}
Our purpose is archived if we show (\ref{argument of RHS}) is less than or equal to (\ref{LHS}) for any $i'_1, i'_2, \dots i'_N, j'_1, j'_2, \dots , j'_N$.  Hereafter we fix $i'_1, i'_2, \dots i'_N, j'_1, j'_2, \dots , j'_N$.   First we compare (\ref{argument of RHS}) and the arguments (\ref{LHS}) associated with $i_l=i'_l$, $j_i=j'_l$ or $i_l=j'_l$, $j_i=i'_l$.  Then, we obtain
\begin{equation}
\begin{aligned}
  & \max_{I, J}\Bigl(\up[C_{i_1i_2\dots i_N}]+\up[C_{j_1j_2\dots j_N}]+ 
\sum_{l=1}^N l p_{i_l}+ \sum_{l=1}^N lp_{j_l}-\sum_{l=1}^{k_2} p_{i_l}-\sum_{l=1}^{k_1}  p_{j_l} +\sum_{l=k_3}^N  p_{j_l}\Bigr)   \\
  -& \left(  \up[C_{i'_1i'_2\dots i'_N}]+\up[C_{j'_1j'_2\dots j'_N}]+ \sum_{l=1}^N l p_{i'_l}+ \sum_{l=1}^N lp_{j'_l}-\sum_{l=1}^{k_1} p_{i'_l}-\sum_{l=1}^{k_2}  p_{j'_l} +\sum_{l=k_3}^N  p_{j'_l}\right)\\
  \ge  &  \max\Bigl(\up[C_{i'_1i'_2\dots i'_N}]+\up[C_{j'_1j'_2\dots j'_N}] + \sum_{l=1}^N l p_{i'_l}+ \sum_{l=1}^N lp_{j'_l}-\sum_{l=1}^{k_2} p_{i'_l}-\sum_{l=1}^{k_1}  p_{j'_l} +\sum_{l=k_3}^N  p_{j'_l},  \\
  &\up[C_{j'_1j'_2\dots j'_N}]+\up[C_{i'_1i'_2\dots i'_N}] + \sum_{l=1}^N l p_{j'_l}+ \sum_{l=1}^N lp_{i'_l}-\sum_{l=1}^{k_2} p_{j'_l}-\sum_{l=1}^{k_1}  p_{i'_l} +\sum_{l=k_3}^N  p_{i'_l}\Bigr) \\
  -&\left( \up[C_{i'_1i'_2\dots i'_N}]+\up[C_{j'_1j'_2\dots j'_N}]+ \sum_{l=1}^N l p_{i'_l}+ \sum_{l=1}^N lp_{j'_l}-\sum_{l=1}^{k_1} p_{i'_l}-\sum_{l=1}^{k_2}  p_{j'_l} +\sum_{l=k_3}^N  p_{j'_l}\right)\\
  =& \max\Bigl(\sum_{l=k_1+1}^{k_2} (-p_{i'_l}+p_{j'_l})  , \quad \sum_{l=k_3}^N  (p_{i'_l} - p_{j'_l})\Bigr).  
\end{aligned}
\end{equation}  
It takes a nonnegative value when 
\begin{equation}  
  \sum_{l=k_1+1}^{k_2} (-p_{i'_l}+p_{j'_l})\ge 0   \quad \text{or} \quad \sum_{l=k_3}^N  (p_{i'_l} - p_{j'_l})\ge 0
\end{equation}
holds  for any $0\le k_1<k_2<k_3\le N$.  Next let us compare (\ref{argument of RHS}) and (\ref{LHS}) in the case    
\begin{equation}  \label{app2-2}
  \sum_{l=k_1+1}^{k_2} (-p_{i'_l}+p_{j'_l})<0  \quad \text{and } \quad \sum_{l=k_3}^N  (p_{i'_l} - p_{j'_l})<0  
\end{equation}
hold for certain $k_1$, $k_2$ and $k_3$.  We introduce the notations 
\begin{equation}  \label{ijset1}
  i'_l = 
\begin{cases}
  1 & \quad (l=1, 2, \dots, \alpha )\\
  2 & \quad (l=\alpha+1, \alpha +2, \dots, \beta)\\
  3 & \quad (l=\beta +1, \beta +2, \dots, N)
\end{cases}, \qquad \qquad 
  j'_l=
\begin{cases}
  1 & \quad (l=1, 2, \dots, \gamma )\\
  2 & \quad (l=\gamma+1, \gamma +2, \dots, \delta)\\
  3 & \quad (l=\delta +1, \delta +2, \dots, N)
\end{cases}.
\end{equation}
Note that $\alpha $, $\beta $, $\gamma $, $\delta $ should be $k_1+1\le \alpha \le k_2<k_3\le \beta \le N$ and $\alpha < \gamma<\delta  <\beta$ in order to hold (\ref{app2-2}).  In addition it should be $i'_{k_2}=i'_{k_3}=2$, $j'_{k_1+1}=1$, $j'_N=3$, $j'_{\alpha }=1$ and $j'_{\beta } =3$  (See Table 1).  For these  $i'_l$ and $j'_l$, we set $i_l$ and $j_l$ as
\begin{equation}  \label{ijset2}
  i_l = 
\begin{cases}
  1 & \quad (l=1, 2, \dots, \alpha+1 )\\
  2 & \quad (l=\alpha+2, \alpha +3, \dots, \beta)\\
  3 & \quad (l=\beta +1, \beta +2, \dots, N)
\end{cases}, \qquad \qquad 
  j_l=
\begin{cases}
  1 & \quad (l=1, 2, \dots, \gamma-1 )\\
  2 & \quad (l=\gamma, \gamma +1, \dots, \delta)\\
  3 & \quad (l=\delta +1, \delta +2, \dots, N)
\end{cases}.
\end{equation}
\begin{center}
\begin{tabular} {|c||c|c|c|c|c|c|c|c|c|c|c|c|c|c|c|c|c|}\hline
 $l$ &  $1$ & $\ldots $  & $\alpha $ & $\alpha+1 $& $\alpha +2$& $\ldots$ & $\gamma-1$ & $\gamma $& $\gamma +1$& $\ldots$& $\delta $ & $\delta +1$ & $\ldots $ & $\beta $ & $\beta+1$ & $\ldots $ & $N$  \\ \hline
 $i'_l$ &  $1$ & $\ldots $  & $1$ & $2$& $2$& $\ldots$ & $2$ & $2$& $2$& $\ldots$ & $2$ & $2$ & $\ldots $ & $2$& $3$ & $\ldots $ & $3$  \\ \hline
 $i_l$ &  $1$ & $\ldots $  & $1$ & $1$& $2$& $\ldots$ & $2$ & $2$& $2$& $\ldots$ & $2$ & $2$ & $\ldots$ & $2$& $3$ & $\ldots $ & $3$  \\ \hline
 $j'_l$ &  $1$ & $\ldots $  & $1 $ & $1$& $1$& $\ldots$ & $1$ & $1$& $2$& $\ldots$ & $2$ & $3$ & $\ldots $& $3$ & $3$ & $\ldots $ & $3$  \\ \hline
 $j_l$ &  $1$ & $\ldots $  & $1 $ & $1$& $1$& $\ldots$ & $1$ & $2$& $2$& $\ldots$ & $2$ & $3$ & $\ldots $ & $3$ & $3$ & $\ldots $ & $3$  \\ \hline
\end{tabular}
 \\
Table 1: The sets of $i_l$, $j_l$, $i'_l$, $j'_l$ for (\ref{ijset1}) and (\ref{ijset2})
\end{center}
Subtracting (\ref{argument of RHS}) associated with (\ref{ijset1}) from the argument of (\ref{LHS}) associated with (\ref{ijset2}), we obtain
\begin{equation}  
\begin{aligned}
  &\up[C_{i_1i_2\dots i_N}]+\up[C_{j_1j_2\dots j_N}]+ 
\sum_{l=1}^N l p_{i_l}+ \sum_{l=1}^N lp_{j_l}-\sum_{l=1}^{k_2} p_{i_l}-\sum_{l=1}^{k_1}  p_{j_l} +\sum_{l=k_3}^N  p_{j_l}\\
-& \left( \up[C_{i'_1i'_2\dots i'_N}]+\up[C_{j'_1j'_2\dots j'_N}]+ \sum_{l=1}^N l p_{i'_l}+ \sum_{l=1}^N lp_{j'_l}-\sum_{l=1}^{k_1} p_{i'_l}-\sum_{l=1}^{k_2}  p_{j'_l} +\sum_{l=k_3}^N  p_{j'_l} \right) \\
 = &\up[C_{i_1i_2\dots i_N}]+\up[C_{j_1j_2\dots j_N}]-\up[C_{i'_1i'_2\dots i'_N}]-\up[C_{j'_1j'_2\dots j'_N}]  \\
 +& (\gamma -\alpha-1) (p_2-p_1) -\sum _{l=k_1+1}^{k_2} (p_{i_l}-p_{j'_l}) +\sum_{l=k_3}^N(p_{j_l}-p_{j'_l}) \\
 \ge &\up[C_{i_1i_2\dots i_N}]+\up[C_{j_1j_2\dots j_N}]-\up[C_{i'_1i'_2\dots i'_N}]-\up[C_{j'_1j'_2\dots j'_N}] \\
  & + (\gamma -\alpha -1)(p_2-p_1)-(\gamma -\alpha -1) (p_2-p_1) +\sum_{l=k_3}^N(p_{j_l}-p_{j'_l}) \\
 = &\up[C_{i_1i_2\dots i_N}]+\up[C_{j_1j_2\dots j_N}]-\up[C_{i'_1i'_2\dots i'_N}]-\up[C_{j'_1j'_2\dots j'_N}]  +\sum_{l=k_3}^N(p_{j_l}-p_{j'_l}) 
\end{aligned}
\end{equation}
It takes a nonnegative value from Proposition \ref{proposition2}.  Therefore (\ref{ineq}) holds.  

\end{document}